\documentclass{svjour3}                     % onecolumn (standard format)
\pdfoutput=1
\usepackage{graphicx}
\usepackage{latexsym}
\begin{document}

\title{A THz spectrometer using band pass filters}


\author{F. Martini,    	  
	\and E. Giovine 
	\and F. Chiarello 
	\and  P. Carelli 
       %etc. 
}


\institute{F. Martini \at
	CNR IFN-Rome via Cineto Romano 42, 00156 Rome Italy\\	
              E. Giovine \at
              CNR IFN-Rome via Cineto Romano 42, 00156 Rome Italy\\
              F. Chiarello \at
              CNR IFN-Rome via Cineto Romano 42, 00156 Rome Italy\\
              P. Carelli \at
              DSFC, Università dell'Aquila, Italy and CNR IFN-Rome via Cineto Romano 42, 00156 Rome Italy 
              \email{Pasquale.Carelli@gmail.com}
        }

\date{}


\maketitle

\begin{abstract}
	We describe a THz spectrometer operating between 1.2 and 10.5 THz, consisting of band pass filters made with metasurfaces. The source is made of 10 W small black body. The detector is  a high sensitivity room temperature pyroelectric sensor. Various techniques  used to prepare samples are described. The spectra obtained are compared with those measured with a Fourier Transformer Infrared Spectrometer on the same samples. The instrument, using commercial technologies available at the present time, can constitute   an  economical alternative to very expensive spectrometers.
	It has already been successfully used, getting precise spectroscopic measures of many inorganic powders.
	
%Insert your abstract here. Include keywords, PACS and mathematical
%subject classification numbers as needed.
\end{abstract}

\section{Introduction}
\label{intro}

 Spectroscopy is a common technique used to distinguish materials. Perception of colors by mean of human eyes is a natural form of spectroscopy.   Three types of cones (retinal cells), sensitive to three different regions of light, perform a trichromatic color vision. The instrument here described applies a similar technique in the Terahertz (THz) region of the electromagnetic spectrum, using band pass filters centered between 1.3 and 10.5 THz. The instrument is the natural evolution of previous apparatus able only to discriminate between materials \cite{Carelli2017}.
 
In review papers, THz is often  considered the range of frequency between 0.1 and 10 THz\cite{Redo-Sanchez2013}. We think that most materials have a completely different behavior between 0.1-1 THz and 1-10 THz. For this reason a region of spectrum, so extended, described with a single name, can be confusing. The main reason for differences in behavior is due to the fact that, while most  materials are almost transparent for frequency lower than 1 THz, the same substances strongly absorb the radiation between 1 and 10 THz; sources  and detectors in the two ranges of frequencies are completely different too. The instrument here described can be modified to be used at frequencies lower than 1 THz, just using a different kind of sources.

The Fourier Transformer Infrared Spectrometer (FTIR)\cite{griffiths2007fourier} is probably the most suitable and used instrument for frequencies higher than 1 THz. In these years the Time Domain Spectroscopy (TDS)  has obtained impressive results and it is successfully employed  for frequency lower than 2 THz\cite{liebermeister2019}, even if there are examples of systems operating at a higher frequency\cite{shen2003}. The FTIR can be used up to the visible part of the spectrum, but it needs vacuum to operate in  THz range; its optical complexity justifies its high cost that inhibits a widespread use. Even  TDS are complex and expensive instruments; on the contrary the instrument here proposed is approximatively two orders of magnitude less expensive than the FTIR and allows to perform spectroscopy in a very simple way, at the expenses of spectral resolution. At room temperature many substances have a very wide absorption region, so a very simple instrument can give all the required information. To show that our instrument is able to perform a real spectroscopy, we compare the results obtained with our instrument with those got by a high quality commercial FTIR on the same samples.

THz spectroscopy is  a useful technique for the study and characterization of many materials: explosives or  drugs \cite{Davies2008}, agri-food  products \cite{WANG2017}, pharmaceutical substances \cite{SHEN201148}, macromolecules of biological interest \cite{YANG2016},  and even pigments used in art \cite{Fukunaga2007}. Pigments used in art are very common, already available in thin powder form, so many of them were used to test this apparatus.

THz spectroscopy provides information on the basic structure of molecules and it is widely used in radio astronomy\cite{siegel2007}. The vibrational frequencies of large molecules and optical phonon of solids fall in this spectral region. Molecules or biological solids have large absorption spectral regions that can be used as fingerprints to discriminate materials. The apparatus described here can be a simple way to perform spectroscopy in this range of values.
Its interest is due to the possibility of having very simple, reliable and cheap instruments for spectroscopic analysis in the frequency range between 1 and 10 THz.

\section{The instrument} 

\begin{figure}[ht]
	\centering
	\includegraphics[width=0.5\linewidth]{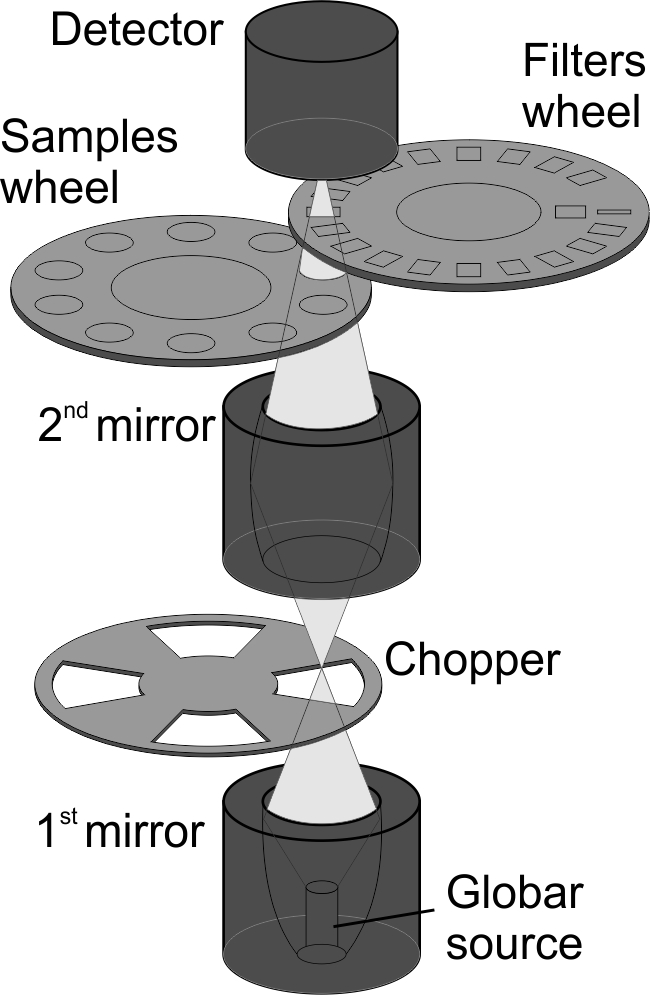}
	\caption{Scheme of the apparatus.}
	\label{fig:soloapparato}
\end{figure}

The physical principles of the instrument are similar to those described elsewhere\cite{Carelli2017}, but there are significant improvements that changed all performances. Figure \ref{fig:soloapparato} illustrates the scheme of the apparatus which allows to explain its functioning, while a photo of the apparatus is shown in Online Resource (fig.1).

 The apparatus works vertically, a small hot black body (globar) is the source placed on the first focus of one ellipsoidal metal on-axis mirror, a second mirror collects the radiation on the sample. The filters and the pyroelectric detector are a few mm above, on the same optical line. The whole optical length is approximately 12 cm. In some measurements we used a third mirror to better focus the signal on the pyroelectric detector: this causes an increase in the total optical length and this addition was made only when necessary as the optical path should be kept as short as possible to reduce the absorption from the atmosphere.

The samples are kept horizontally, so that the study of many kinds of materials is simplified: we can analyze solid slices, powders, soft materials and liquids. Here we present just measurements on some powders. The whole instrument has an height of 45 cm and a base 20x20 cm, so that it can be easily put in a small box with a controlled atmosphere, reducing  humidity to negligible values, if necessary. The weight of the apparatus is only of a few kilograms.

The system can also be easily assembled in a different way to perform $45^o$ reflection measurements by adding a third mirror integrated with the pyroelectric detector.  The additional mirror has the first focus on the surface of the reflecting material, while the second focus is on the detector in order to collect the radiation. The filters are between the third mirror and  the pyroelectric. This is not a true reflection measurement because the metal support is the real reflecting surface, due to the negligible thickness of the material. In this case the total optical lenght is increased to a value of 17cm.

The source is a small black body (globar \cite{globar}) made with a 1.4 mm diameter silicon carbide rod. The carbide rod is an igniter requesting just a 6 V supply. It is kept at 1200 K with a power supply of only 10 W. A small source and a low power guarantee greater durability and simplify the whole system, so that it is not necessary to remove the excess heat; in addition, local heating has a beneficial effect on the measurements, locally creating a zone of lower humidity.

The metallic mirrors are sandblasted and covered with black carbon to reduce  the infrared background \cite{Carelli2017} and are similar to those of the previous apparatus, but the new design allows an easy alignment along the optical axis.

A chopper in the focus, shared between the two mirrors, modulates the radiation at a frequency of 10 Hz, a value  chosen in order to maximize the signal-to-noise ratio.  The reference signal from the chopper is used for an accurate lock-in analysis made via software.

We used various sample holders. The ones used for the samples compared with the measurements with the FTIR consists of a polycarbonate disc (a simple CD) with ten 12 mm holes on a 46 mm radius circle. The HDPE samples, described later, are placed on the holes. The rotation of the sample holder is controlled by a stepper motor. Another stepper motor, fixed to the box containing the pyroelectric detector, rotates the metal support of the silicon wafer (the substrate for the THz filters). A LabVIEW-based computer software positions the two stepper motors through an Arduino board, controlling the sequence of the filters and of the samples to be analyzed. 

The metallic mirrors are fabricated by a CNC (Computer Numerical Control)  machine and their cost is at least an order of magnitude less than glass mirrors. We use a high quality chopper wheel (model M1F2 by Thorlabs), but with a simple dc-motor. The trigger signal is acquired by common low cost devices: IR-LED and IR receiver. 

The alignment of the optics for the previous apparatus \cite{Carelli2017} was a difficult task: the assembly of the instrument and the alignment required weeks of work, due to the six degrees of freedom of each mirror. The new  mechanical project allows us to leave just a degree of freedom for each mirror: a movement along the common optical axis. Aligning  the sample holder, filters and pyroelectric detector  along the central line of the mirrors is an easy job. In one day all the apparatus can be mounted or demounted, thanks also to the THz wavelength that needs a mechanical precision of about $10\ \mu m$. 

The used detector is the same of the one described in the previous paper\cite{Carelli2017}, a commercial pyroelectric sensor (QS2-THz-BL Gentec-EO), but we have
substituted  the factory silicon window with a thick layer of black HDPE; this  has given a significant improvement to the apparatus. This new window has been characterized in the range 1.5 and 17 THz by means of a FTIR and its transmission curve in THz range is well described by a simple quadratic polynomial:
$$T(\nu)=0.95-0.073\nu+0.0018\nu^2$$
where $\nu$ is the frequency in THz.
This window reduces the amplitude of the signal collected at 10 THz of about the 60\%, but  more significantly it reduces the residual mid-IR spurious signal,  that is not eliminated by the absorption of the ellipsoidal mirrors.

\section{Selective filters} 

\begin{figure}[ht]
	\centering
	\includegraphics[width=1\linewidth]{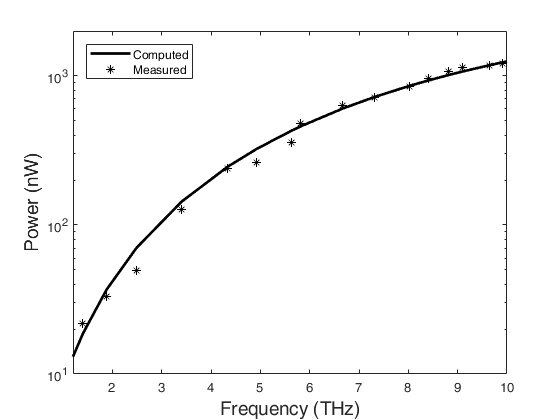}
	\caption{Expected and measured performances of the eighteen band pass filters. }
	\label{fig:performances}
\end{figure}

The selective filters are realized by means of metasurfaces microfabricated on a silicon substrate \cite{Carelli2017}.  Metasurfaces are periodic 2D structures of  complementary rings (annular holes on the metallic film). The filter with a central frequency of 10.5 THz has a square cell of lattice pitch $ 7 \mu m $, with a ring in the middle of outer diameter $ 4.9 \mu m $ and a groove of $ 0.75 \mu m $, while the lower frequency filter (1.2 THz) has a lattice pitch of $ 56 \mu m $, an outer diameter of $ 39 \mu m $ and a groove of $ 3 \mu m $.

We span frequencies higher than those investigated by the previous apparatus: the highest filter has a central frequency  of 10.5 THz. Only three filters operate between 1 and 2 THz, so most of the filters operate in a higher frequency range, where the power of the source is higher. 

The manufacturing process is based on a direct writing with electron beam lithography on a 75 mm double-side polished silicon wafer, previously covered with a layer of about 200 nm of aluminum and an electronic resist. Once developed the film is  wet etched. For our process the wet etch  is appropriate for low frequency filters, but it is a technological limit for those at higher frequency.

The 18 filters have been characterized by a FTIR and their transmitted signal has been well fitted by a Lorentz distribution:
\begin{equation}\label{filter}
T(\nu)=A \frac {(\nu_o/2Q)^2 }{(\nu-\nu_o)^2+(\nu_o/2Q)^2}
\end{equation}
where $\nu$ is the frequency in $THz$, $\nu_0$ is the central frequency, $Q=3.5$ and  $A$ is the amplitude of the maximum.  $\nu_0$ is a linear function of the project scaling factor: we have a base structure at 3.7 THz and to obtain the highest frequency filter (10.5 THz) we multiply all measures for 0.35 (the scaling factor),  and for  2.8 to obtain that of the lowest frequency (1.2 THz).

The use of  the Lorentz distribution  for computing the expected signals is a good approximation, even if it does not take into account the residual mid-IR spurious signal at higher frequency; this signal cannot be easily measured, but just attenuated by the  window of  black HDPE on the detector and by the ellipsoidal mirrors.

In fig. \ref{fig:performances} we plot the expected signal by each filter (continuous line) and the experimental results (stars). For the expected signal we use the Lorentz distribution of filters, the HDPE window transmission curve and the emission of the black body source. The ratio  between the radiation passing through the higher  (10.5 THz) and the lower (1.2 THz) filters is expected to  be approximatively one hundred and it is experimentally verified. The black HDPE window and the metal mirrors are both necessary in order to reduce the spurious signal in the mid-infrared. 

\section{Main sample preparation}
The transmission characteristic curve of the material is obtained by the ratio between the signal measured by each filter with the sample and the one obtained with an empty disc of HDPE. The measurement error depends on the amplitude of the signal: if we have a power greater than 50 nW the  error is less than 1 \% (with an acquisition time of 10 s for each filter); if the power is lower than 5 nW, the  error can be larger than 10 \%. A complete measurement (with 18 filters) takes about 200 s, while a fast measurement (1 s for filter) can be performed in 36 s. To avoid confusion, from now on we'll call the measuring apparatus with  an acronym  \textit{MFA} (Metasurface Filters Apparatus).

We tested  many powdered inorganic materials with the \textit{MFA}:  $Al(OH)_3$, $Al_2O_3$, $BaCO_3$, $CaCO_3$, $CaSiO_3$, $CuO$, $FeO$, $KBr$, $Li_2CO_3$, 
$Mg(OH)_2$, 
$MnO_2$, $Na_2CO_3$, $NaCl$, $NaF$, $SiO_2$, $SnO_2$, $ZnO$, $ZrSiO_4$. Most of them are fine powders used as ceramic pigments. To avoid effect due to the dispersion of radiation the maximum size limit for the grains is about $20\ \mu m$. The size of grain of ceramic pigments is always lower that this limit.  The others materials are soft and can be easily ground to reduce their grain size, if necessary. 

All these materials are stable over a wide range of temperatures. To prepare the samples we distribute the powders to be investigated  on an HDPE disc (10 mm initial diameter and 1 mm thickness), compress the stack between two glass slides and put it in an oven at $140 ^oC$ for a few minutes. The procedure causes a thinning and a flattening of the disc, but the powder doesn't change its position. HDPE is a thermoplastic polymer that easily supports the thermal cycle and is almost transparent to THz. We usually put more powder  than needed. The disc is measured with a rapid measurement, the excess material is removed and the disc is covered with a thin layer of HDPE ($ 10\ \mu m$ thick) and placed back into the oven. So we have stable samples that can be measured by the \textit{MFA} and FTIR. The protective layer prevents problems of hygroscopicity and guarantees the stability of the samples.  After we measure the weight of the samples we can estimate their average thickness. The two faces of the samples are intentionally not parallel (when possible) in order to remove the artifact in the FTIR (signal modulation)  caused by the interference between them. The flatness of the surface guarantees a higher signal in both the used instruments .

The sample preparation takes a few minutes and we have prepared two samples for all the materials at least:
 some of them ($BaCO_3$, $CaCO_3$, $CaSiO_3$, $KBr$, $Li_2CO_3$, $NaF$ and $SnO_2$) show clear features in the THz spectrum with the \textit{MFA}; the other ones, even if they are easily discriminated, have less interesting features.

\section{Experimental validation}
\begin{figure}[ht]
	\centering
	\includegraphics[width=1\linewidth]{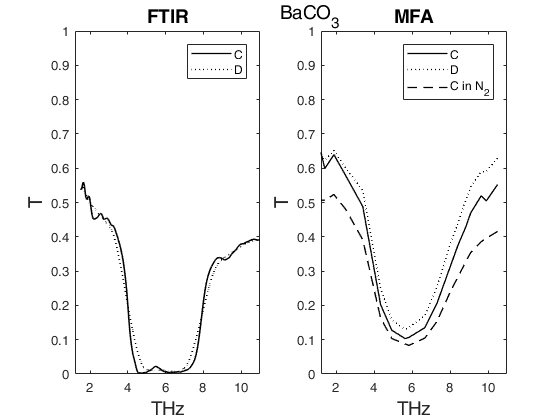}
	\caption{Comparison on two samples  C and D of $BaCO_3$ using FTIR (left) and \textit{MFA} (right).}
	\label{fig:baco3fiftir}
\end{figure}

To validate the \textit{MFA} we performed transmission measurement of the same sample with the FTIR.

The first measurements presented here  is on  Barium Carbonate ($BaCO_3$) as shown in fig.\ref{fig:baco3fiftir}.

 The two samples have almost the same equivalent thickness of about $22\ \mu m$.
 The two measurements give similar results.  To be noted, the filters at lower frequency are affected by an error that can be also of 10\%, because of the small signal, then the measure in this range of frequency is often less significant. In the \ref{fig:baco3fiftir} the sample C was additionally  measured in pure Nitrogen atmosphere, finding a smoother characteristic, but no significant differences. The FTIR measurement is made using vacuum as reference (not a flat disk of HDPE); for this reason the phonon peak of polyethylene at 2.2 THz appears.
 \begin{figure}[ht]
 	\centering
 	\includegraphics[width=1\linewidth]{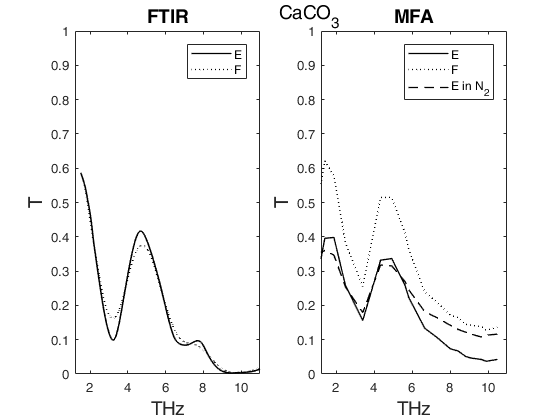}
 	\caption{Comparison on two samples C and D of $CaCO_3$ using FTIR (left) or filter (right).}
 	\label{fig:caco3fiftir}
 \end{figure} 
 
The second measurement here presented is on Calcium Carbonate ($CaCO_3$) and 
is shown in fig. \ref{fig:caco3fiftir}.

\begin{figure}[ht]
	\centering
	\includegraphics[width=1\linewidth]{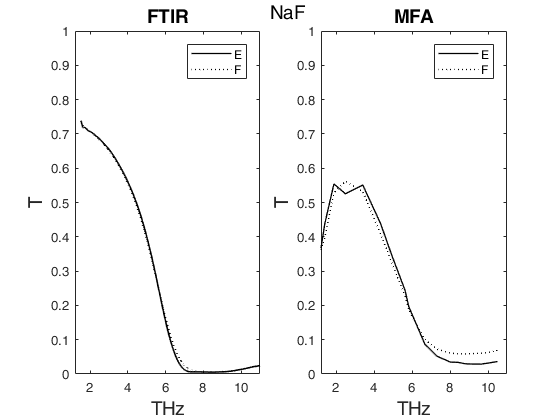}
	\caption{Comparison on two samples E and F of $NaF$ using FTIR (left) or filter (right).}
	\label{fig:naffiftir}
\end{figure}

Two samples were measured: sample E $19\ \mu m$ and sample F  $10\ \mu m$. The \textit{MFA} correctly measures a different amplitude correlated with the thickness of the material. The absence of the peak at 7.5 THz with the \textit{MFA} measurements  is due to the fact that its bandwidth  is too wide to discriminate small features.
Also in this case we made a measurement in pure Nitrogen: here there is a more significant difference between 8 and 10 THz.

The third measurement is performed on an alkaline halides: the sodium fluoride ($NaF$), shown in fig. \ref{fig:naffiftir} .

The two samples have different thickness:  sample E $39\ \mu m$
 and  sample F $21\ \mu m$.  Our results confirm  the trasmittance of alkaline halides  known in literature (see e.g. pag. 60 of the review book \cite{brundermann2012terahertz}).

The measures with MFA were made using a flat disk of HDPE as a reference, while in those with
FTIR the vacuum was used as reference, in order to reduce the possibility to have  interference
between the two faces of the samples. The measurement on the all other materials are available on Online Resource (fig.2 $\cdots$ fig.16).

\section{Other ways of sample preparation}

\begin{figure}[ht]
	\centering
	\includegraphics[width=1\linewidth]{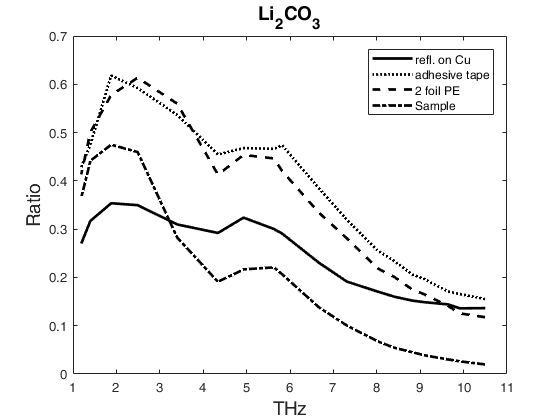}
	\caption{The \textit{MFA} used to test different typologies of sample preparation for $Li_2CO_3$: Solid line: the material is on a flat copper surface. Dotted line: the powder is on adhesive tape. Dashed line: the powder is between two thin sheet of polyethylene. Dashed-dot line: the powder is inglobate in HDPE (the stable sample preparation).}
	\label{fig:lico3varie}
\end{figure}
 
 We tried other preparation techniques that give the same spectrum for many materials, like those shown in  fig.\ref{fig:lico3varie}   for Lithium Carbonate ($Li_2CO_3$): 
 \begin{enumerate}
 	\item[$\bullet$]  continuous line:  The pigment dispersed in a small amount of water is distributed on a smooth surface of copper with a brush and left to dry. The radiation has an angle of incidence at $45^o$ on the sample, the ray reflected at the same angle is collected by the pyroelectric with an additional ellipsoidal mirror.  The ratio between this measurement and that one obtained with the glossy copper is the shown spectrum.  We performed measurements using a thicker layer of material, but the signal was too low to obtain meaningful information.
 \end{enumerate}	
 All the other measurements are made with the instrument with two mirrors, as described in fig. \ref{fig:soloapparato}. 
 \begin{enumerate}
 	\item[$\bullet$] dotted line: We used the common adhesive tape, that is usually almost transparent at THz, as support and simply brushed some powder on the glue of the tape.
 	\item[$\bullet$]  dashed line: The samples are distributed between two very thin foils of polyethylene (less than $20\ \mu m$ each) and sealed with an o-ring on a large rigid plastic support.
 	\item[$\bullet$] dashed dotted: the stable sample preparation ( powder inglobate in HDPE). 
 \end{enumerate}	
 
 The last two methods can be both used to compare samples with the FTIR. But using a sample between two very thin foils of polyetilene, the parallelism of the sample produces an artifact (modulation) on the FTIR measurements.
 All the four techniques are suitable for spectroscopy, even if we normally use the most stable one with the disk of HDPE. This method can be used for materials  stable for a temperature below $140 ^oC$. The use of adhesive tape to prepare samples, is the simplest technique, but it is more sensitive to dust during preparation.

\section{Conclusions}
In this paper we show that the use of selective filters can be an easy way to perform spectroscopy in the THz range of frequencies. The system described here does not require a special preparation of the samples, it works in air and can discriminate features in the limit of the bandwidth of the filters used. Actually we can discriminate features covering a range larger than $\nu _o/3.5$, what shown in 
fig. \ref{fig:caco3fiftir} has been found for other materials.

The choice of a low Q has been made in order to have a very large range of frequencies to investigate, because the choice of materials was absolutely casual, as we did not have any previous information. 

There is not a technological problem to implement band-pass filters with a Q factor that is many times larger than the one used for this work (we can easely increase the Q reducing the shape of the element of the metasurface) and consequently increasing the spectral resolution. We think that our \textit{MFA} could be used to discriminate different kinds of interesting materials like explosives (see e.g. \cite {leahy2007}) or drugs (see e.g \cite{Davies2008}). 

Most materials have a large absorption coefficient in THz range of frequencies: a few tens of milligrams are usually enough to analyze a substance. This allows to  analyze extremely limited quantities, an advantage on the study  of art-works pigments.

The sample preparation is much easier in THz, if compared with the mid-infrared range;
the larger wave length allows less constrains. We showed that many simple techniques of sample preparation can be successfully used.
This instrument is a low cost alternative to standard THz spectrometer, its interest is in its extreme simplicity and low cost, suitable for a widespread use.

\section{Acknowledgment}

The present work was partially supported by the University of l'Aquila (Fondi premiali Carelli) during 2017. We thank G. Torrioli for many  electronic suggestion. We thank F. Del Grande for his mechanical support. We thank the Center of Microscopy of L'Aquila University which gave us the possibility to use the FTIR. F.M. acknowledges the support of the H2020 Marie Skłodowska-Curie Actions (SHAMROCK, 795923).

%\section*{Conflict of interest}
%
%The authors declare that they have no conflict of interest.

\medskip

%\bibliographystyle{unsrt}      %entries are in order of citation. 
\bibliographystyle{spphys}       % APS-like style for physics
\bibliography{ms}   % name your BibTeX data base

\begin{thebibliography}{10}
\providecommand{\url}[1]{{#1}}
\providecommand{\urlprefix}{URL }
\expandafter\ifx\csname urlstyle\endcsname\relax
  \providecommand{\doi}[1]{DOI \discretionary{}{}{}#1}\else
  \providecommand{\doi}{DOI \discretionary{}{}{}\begingroup
  \urlstyle{rm}\Url}\fi

\bibitem{Carelli2017}
P.~Carelli, F.~Chiarello, G.~Torrioli, M.G. Castellano, Journal of Infrared,
  Millimeter, and Terahertz Waves \textbf{38}, 303 (2017)

\bibitem{Redo-Sanchez2013}
A.~Redo-Sanchez, N.~Laman, B.~Schulkin, T.~Tongue, Journal of Infrared,
  Millimeter, and Terahertz Waves \textbf{34}, 500 (2013)

\bibitem{griffiths2007fourier}
P.~Griffiths, J.~De~Haseth, J.~Winefordner, \emph{Fourier Transform Infrared
  Spectrometry}.
\newblock Chemical Analysis: A Series of Monographs on Analytical Chemistry and
  Its Applications (Wiley, 2007)

\bibitem{liebermeister2019}
L.~Liebermeister, S.~Nellen, R.~Kohlhaas, S.~Breuer, M.~Schell, B.~Globisch,
  Journal of Infrared, Millimeter, and Terahertz Waves \textbf{40}(3), 288
  (2019)

\bibitem{shen2003}
Y.~Shen, P.~Upadhya, E.~Linfield, H.~Beere, A.~Davies, Applied physics letters
  \textbf{83}(15), 3117 (2003)

\bibitem{Davies2008}
A.G. Davies, A.D. Burnett, W.~Fan, E.H. Linfield, J.E. Cunningham, Materials
  today \textbf{11}(3), 18 (2008)

\bibitem{WANG2017}
K.~Wang, D.W. Sun, H.~Pu, Trends in Food Science \& Technology \textbf{67}, 93
  (2017)

\bibitem{SHEN201148}
Y.C. Shen, International Journal of Pharmaceutics \textbf{417}(1), 48  (2011).
\newblock Advanced characterization techniques

\bibitem{YANG2016}
X.~Yang, X.~Zhao, K.~Yang, Y.~Liu, Y.~Liu, W.~Fu, Y.~Luo, Trends in
  Biotechnology \textbf{34}(10), 810  (2016)

\bibitem{Fukunaga2007}
K.~Fukunaga, Y.~Ogawa, S.~Hayashi, I.~Hosako, IEICE Electronics Express
  \textbf{4}(8), 258 (2007)

\bibitem{siegel2007}
P.H. Siegel, IEEE Transactions on Antennas and Propagation \textbf{55}(11),
  2957 (2007)

\bibitem{globar}
\emph{ML6-10-203FS by Crystal Tecnica Ltd UK}

\bibitem{brundermann2012terahertz}
M.F.K. Erik~Br{\"u}ndermann, Heinz-Wilhelm~H{\"u}bers, \emph{Terahertz
  techniques}, vol. 151 (Springer, 2012)

\bibitem{leahy2007}
M.~Leahy-Hoppa, M.~Fitch, X.~Zheng, L.~Hayden, R.~Osiander, Chemical Physics
  Letters \textbf{434}(4-6), 227 (2007)

\end{thebibliography}

\end{document}

\title{ELECTRONIC SUPPLEMENTARY MATERIAL \\
	A THz spectrometer using band pass filters}


\author{F. Martini,    	  
	\and E. Giovine 
	\and F. Chiarello 
	\and  P. Carelli }

\date{}

\maketitle

\newpage

\begin{figure}[ht]
	\centering
	\includegraphics[width=1\linewidth]{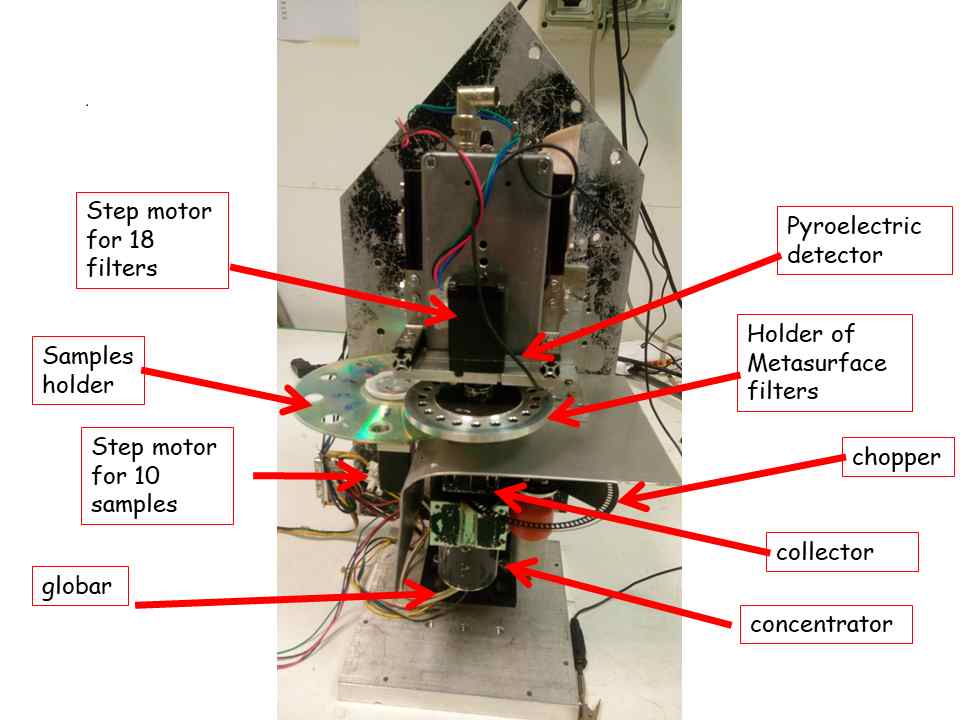}
	\caption{Photo of the apparatus.}
	\label{fig:soloapparato}
\end{figure}

\noindent
In fig.\ref{fig:soloapparato} there is a photo of the apparatus. The globar (black body source) is inside the concentrator (first mirror). The second mirror and the chopper are almost hidden by a shield (not described in the text). The sample holder in the picture has no samples. The step motor that rotates the 18 filters hides the window of the pyroelectric detector. The box supporting the step motor contains the battery power supply for the pyroelectric detector.

\newpage

\noindent
In fig.2 $\cdots$ fig.17 a comparison between the samples with FTIR (left) and metasurface filters apparatus (MFA, right) is shown. Some of the measures (Calcium silicate, Sodium carbonate, Zinc oxide) with FTIR have an artifact due to the interference between the two faces of the samples: in these cases we have smoothed the results. For Lithium carbonate, even if needed, we didn't smooth data to give an idea of the effect of non-smoothing.
\vspace{1cm}

\noindent%
%\begin{minipage}{\linewidth}% to keep image and caption on one page
	%\makebox[\linewidth]{%        to center the image
		\includegraphics[keepaspectratio=true,scale=0.8]{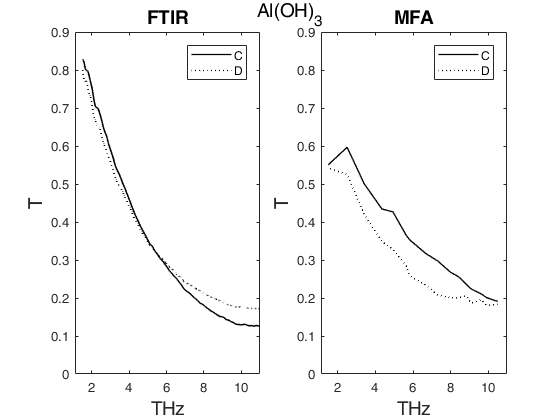}
	\captionof{figure}{Aluminium hydroxide}
%\end{minipage}

\begin{figure}[h]
	\centering
	\includegraphics[width=0.95\linewidth]{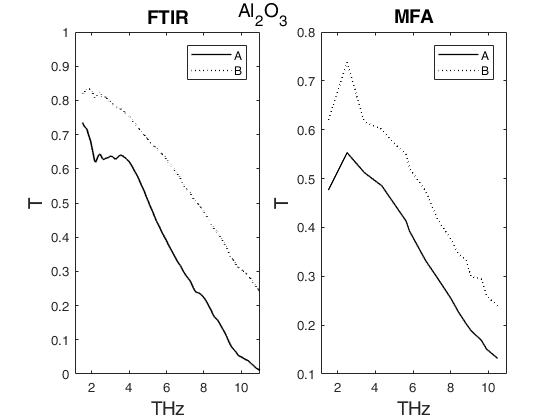}
	\caption{Aluminum oxide}		
	\vspace{1cm}
	\includegraphics[width=0.95\linewidth]{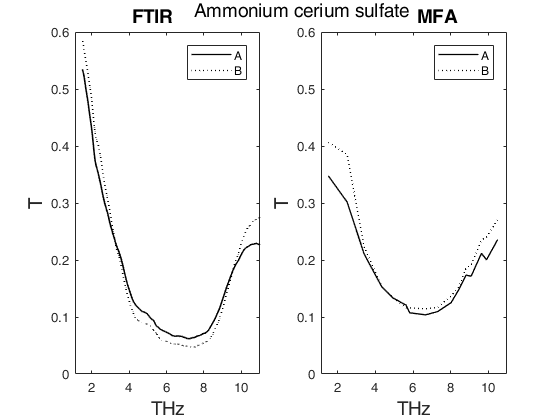}
	\caption{Ammonium cerium(IV) sulfate}			
\end{figure}

\begin{figure}[h]
	\centering
 	\includegraphics[width=0.95\linewidth]{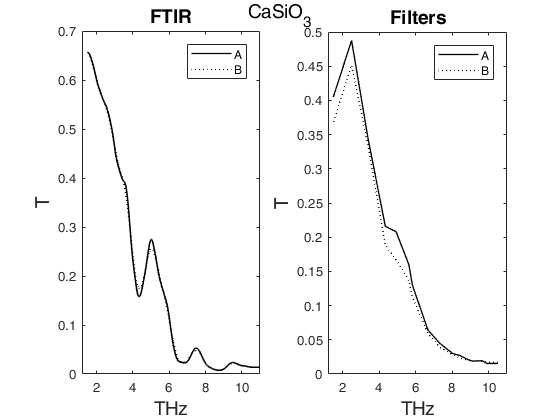}
	\caption{Calcium silicate}		
	\vspace{1cm}
	\includegraphics[width=0.95\linewidth]{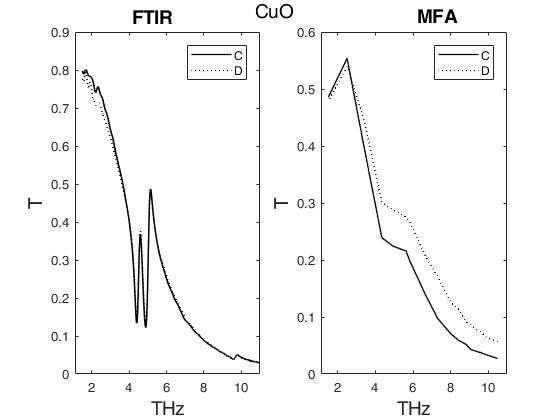}
\caption{Copper oxide}			
\end{figure}

\begin{figure}[h]
	\centering
	\includegraphics[width=0.95\linewidth]{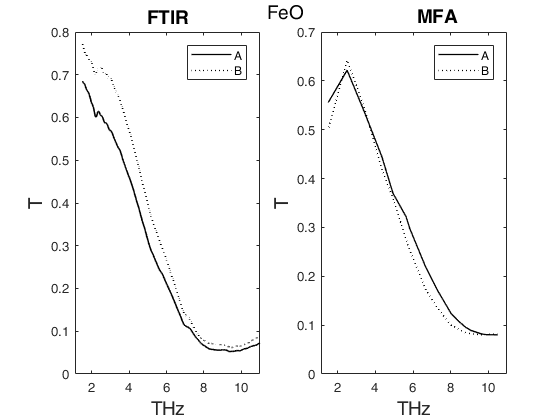}
\caption{Iron(II) oxide}					
	\vspace{1cm}
	\includegraphics[width=0.95\linewidth]{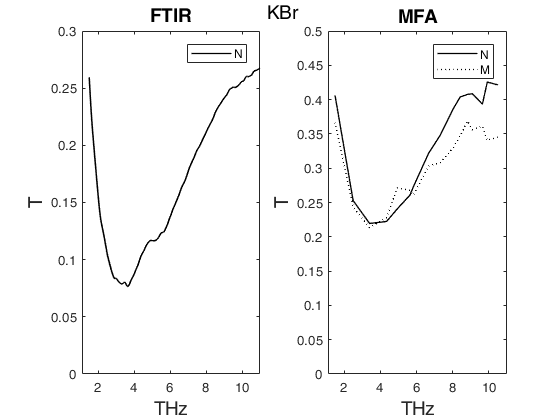}
\caption{Potassium bromide}
	
\end{figure}

\begin{figure}[h]
	\centering
\includegraphics[width=0.95\linewidth]{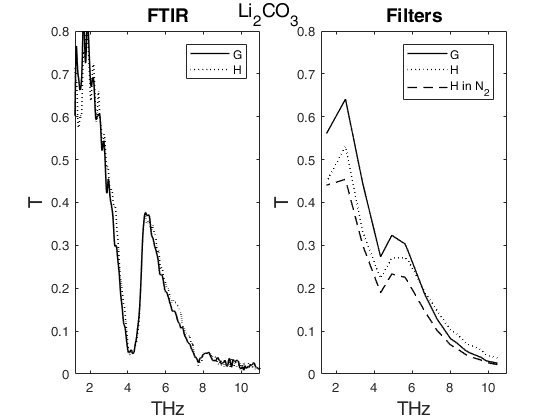}
\caption{Lithium carbonate}
	\vspace{1cm}	
	\includegraphics[width=0.95\linewidth]{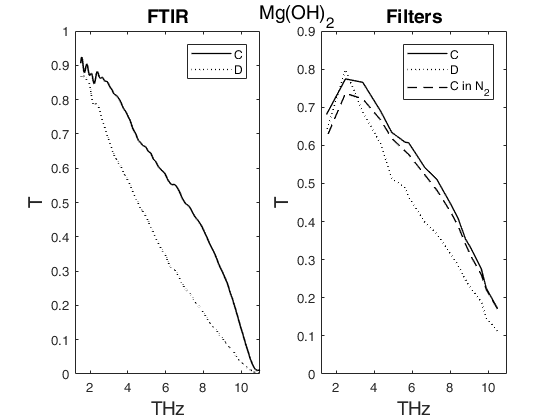}
\caption{Magnesium hydroxide}
\end{figure}

\begin{figure}[h]
	\centering
	\includegraphics[width=0.95\linewidth]{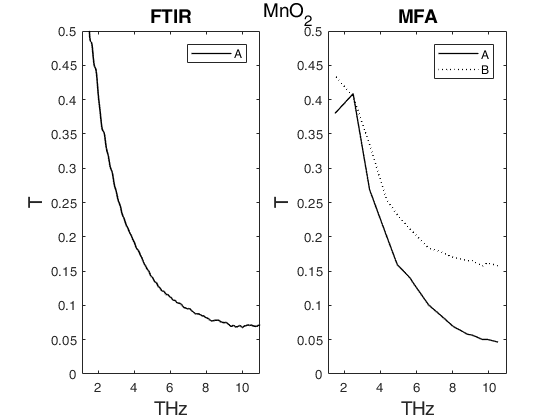}
\caption{Manganese(IV) oxide}	
	\vspace{1cm}
\includegraphics[width=0.95\linewidth]{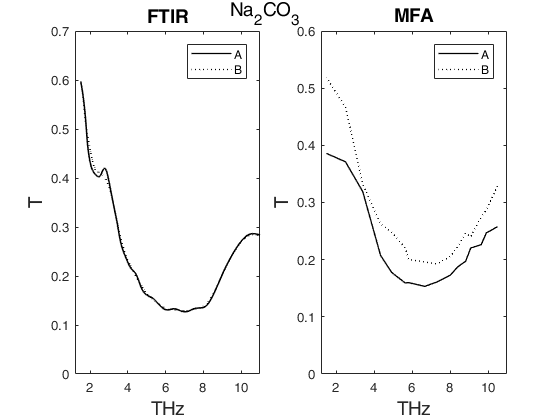}
\caption{Sodium carbonate}
\end{figure}

\begin{figure}[h]
	\centering
	\includegraphics[width=0.95\linewidth]{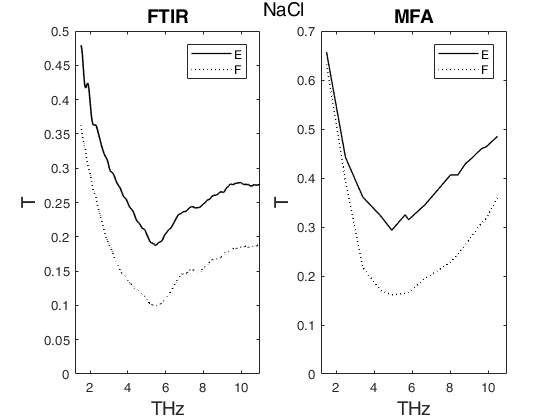}
	\caption{Sodium chloride}
	\vspace{1cm}
	\includegraphics[width=0.95\linewidth]{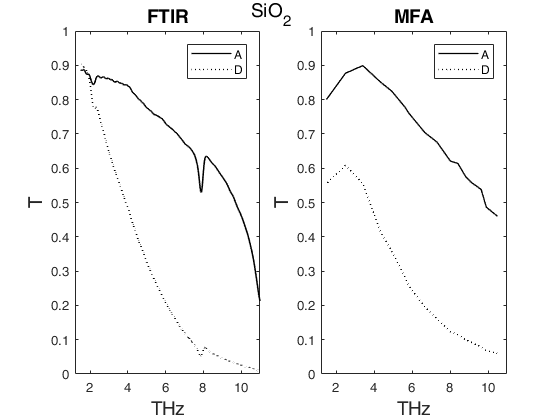}
\caption{Silicon dioxide }
\end{figure}

\begin{figure}[h]
	\centering
	
	\includegraphics[width=0.95\linewidth]{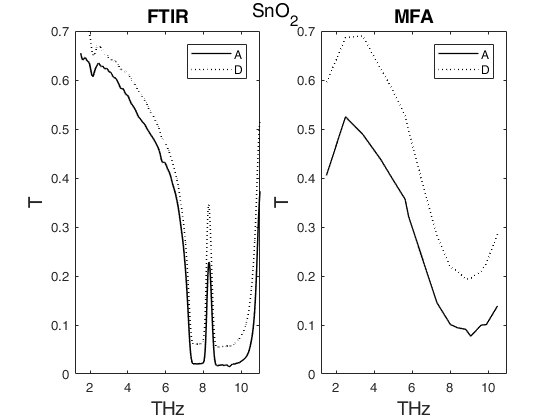}
	\caption{Tin(IV) oxide}
	
	\vspace{1cm}
	
	\includegraphics[width=0.95\linewidth]{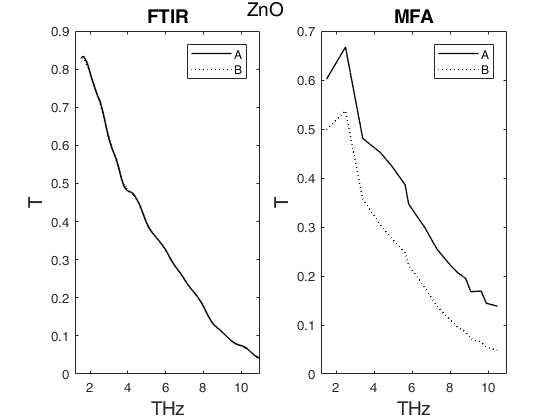}
	\caption{Zinc oxide }
	
\end{figure}

\begin{figure}[h]
	\centering
	
	\includegraphics[width=0.95\linewidth]{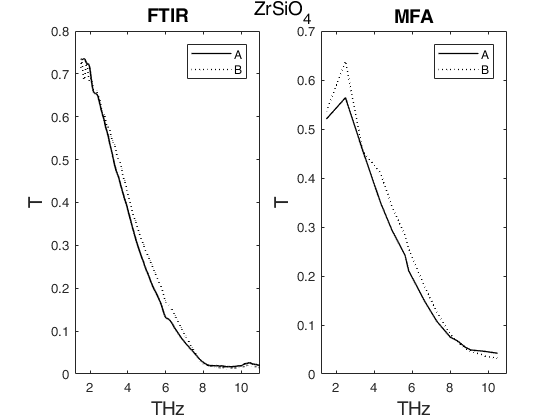}
	\caption{Zirconium silicate}

\end{figure}